\documentclass[preprint,pre,aps,superscriptaddress,a4paper]{revtex4}

\usepackage[utf8x]{inputenc}
\usepackage{graphicx, 
}
\usepackage{epstopdf}
\usepackage{amsmath}
\usepackage{amsfonts}
\usepackage{color}
\begin{document}
\title{Density functional theory for systems with  mesoscopic inhomogeneities}
\author{ A. Ciach}
\affiliation{Institute of Physical Chemistry,
 Polish Academy of Sciences, 01-224 Warszawa, Poland}
 \author{ W. T. Gozdz}
\affiliation{Institute of Physical Chemistry,
 Polish Academy of Sciences, 01-224 Warszawa, Poland}
 \date{\today} 

\begin{abstract}
We study effects of   fluctuations on the  mesoscopic length-scale  on systems with  mesoscopic inhomogeneities.
Equations for the correlation function and for the average volume fraction are derived in the self-consistent Gaussian approximation.
The equations are further simplified by postulating the expression for the structure factor consistent with scattering 
experiments for  self-assembling systems.
Predictions of the approximate theory are verified by a comparison with the exact results obtained earlier for the one-dimensional
lattice model
with first-neighbour attraction and third-neighbour repulsion. We find qualitative agreement for the correlation function, 
the equation of state and the dependence of the chemical potential $\mu$ on the volume fraction $\zeta$. 
Our results confirm also that strong inhomogeneities in the disordered phase are
found only  in the 
 case of strong repulsion. The inhomogeneities are reflected in an oscillatory decay of the correlation function with
a very large correlation length, three inflection points in 
the $\mu(\zeta)$ curve and a compressibility that for increasing $\zeta$ takes very large, very small and again very large values. 

\end{abstract}

\maketitle
\section{Introduction}

Density functional theory (DFT)\cite{evans:79:0} is very successful in describing microscopic properties of 
fluids.
However, when fluctuations at  length scales  
much larger than the molecular size dominate, predictions of the standard DFT are much less accurate. 
Fluctuations at mesoscopic length scales are present in a critical region~\cite{goldenfeld:92:0}, 
at an interface between coexisting fluid phases~\cite{evans:90:2}
and in systems
with competing interactions. Competing interactions are quite
common in soft matter systems~\cite{seul:95:0,sear:99:0,pini:00:0,pini:06:0,imperio:04:0}. For example, charged particles  repel each other at large separations,
but at short separations attract each other with solvent-mediated forces.
The above so called SALR potential  may lead to
self-assembly into clusters  of spherical or elongated shape, into layers 
or into networks, and to a periodic distribution of these objects in space~\cite{archer:07:1,stradner:04:0,ciach:10:1,candia:06:0}. 
Competing interactions are also present in systems containing amphiphilic molecules~\cite{ciach:01:2,ciach:13:0}, 
and in thin magnetic films  with dipolar interactions~\cite{BaSt2007,BaSt2009}.

For systems with competing interactions mean-field (MF) theories, including a local DFT, 
predict stability of several ordered phases 
inside a bell-shaped region in the density-temperature phase 
diagram~\cite{ciach:08:1,ciach:10:1,pekalski:13:0,pekalski:14:0,archer:08:0}. The disordered phase stable outside this region
coexists with two ordered phases only - with periodically
distributed clusters  at low- and   with periodically distributed 
voids at  high density. 
The coexistence line between the disordered and the ordered phases lies close to the line of instability of the disordered phase
with respect to periodic density fluctuations.
When the instability line is approached, both the correlation length and the amplitude 
of the correlation function diverge~\cite{ciach:08:1,ciach:03:1}.
The latter result is clearly nonphysical, since the correlation function for dimensionless density is bounded from above by $1$.

In simulations  the ordered phases have been observed
too, but the phase  diagram is significantly different~\cite{almarza:14:0,imperio:06:0,candia:06:0}. 
The stability region of  the ordered phases
is much smaller~\cite{almarza:14:0,imperio:06:0,candia:06:0}. In addition, 
a less ordered phase is stable between the stability regions of two
different ordered phases~\cite{almarza:14:0,imperio:06:0}, in contrast to the MF results~\cite{ciach:08:1,ciach:10:1,archer:08:0}.
In the case of a triangular lattice model with competing first-neighbor attraction
and third-neighbor repulsion~\cite{pekalski:14:0},  at low temperature $T$ a
molten lamella phase occupies some phase-space
region between the stability regions of the 
  phase with periodically distributed clusters and the 
  phase with periodically distributed stripes~\cite{almarza:14:0}.  At higher $T$ the 
  disordered fluid is stable between the ordered cluster phase and the molten lamella.
  The transition between the fluid and the molten lamella is
  continuous at high $T$ and becomes first order at the tricritical point.
In the molten lamella  discovered in Ref.\cite{almarza:14:0} 
the orientational order of stripes is present but the  translational order is absent.
 A similar  phase  called a ``nematic phase'' was recently detected
in magnetic films with competing interactions~\cite{barci:13:0} in a continuous Brazovskii-type model~\cite{brazovskii:75:0}. 
According to these new discoveries,  instead of the
fluctuation-induced first-order phase transition between the disordered and the periodically ordered phases
predicted earlier 
by Brazovskii~\cite{brazovskii:75:0}, 
a continuous or a first-order transition between the  fluid and the  molten-lamella phases occurs. 

In addition to the presence of new phases and significant modifications of the  phase-coexistence, the 
fluctuations at the mesoscopic length scale lead to significantly different
  properties of the disordered inhomogeneous phase.
In simulation snapshots of the disordered phase clusters or layers are clearly visible, indicating that this phase can
 be very inhomogeneous~\cite{imperio:04:0,imperio:06:0,almarza:14:0, archer:07:1}.
 Even though the distribution of these objects is not 
 periodic, they can be correlated over large distances. The disordered inhomogeneous phase may thus be
 considered as a molten periodic phase, with the long-range order destroyed by the mesoscopic fluctuations.

 The effects of fluctuations on the disordered phase were determined in a one-dimensional (1d) lattice model with
competing first-neighbor attraction and third-neighbor repulsion by the exact transfer matrix method
\cite{pekalski:13:0}.  In the ground state ($T=0$) of this model
a dilute phase (vacuum) is stable for small values of the chemical potential $\mu$
and a dense  phase (fully occupied lattice) is stable for large values of  $\mu$.
When the repulsion is sufficiently strong, an ordered phase with periodic distribution of clusters and density $\rho=1/2$ is stable
for intermediate values of  $\mu$.
At $T>0$ only a disordered phase is stable, but the presence of clusters at low $T$ is reflected in a very large
correlation length, and in a specific shape of  the $\rho(\mu)$ curve. It contains three inflection points at   low $T$ 
 - the central one at $\rho=1/2$ and the  other two
at $\mu$ corresponding to the $T=0$ phase transitions. 
For $T\to 0$  a step-like $\rho(\mu)$  curve was obtained.  
The very small compressibility of the system 
for $\rho\approx 1/2$ indicates formation of clusters that repel each other at short separations. The very large compressibility 
at the other two inflection points signal an
approach to the phase transitions  that occur at $T=0$. 
Only at high $T$  a single
inflection point at $\rho=1/2$ is present.
In contrast, in MF the $\rho(\mu)$ curve has a single inflection point for the whole range of stability
of the disordered phase.
Thus, MF fails to predict the qualitative features of the disordered inhomogeneous phase.
 Only for high $T$, where 
 strong inhomogeneities are no longer present,  qualitative properties of the disordered phase are correctly described by MF.  
 MF predicts stability of a periodically ordered phase instead of strongly inhomogeneous disordered phase.
 
From the comparison of the MF and simulation or exact results it evidently follows that an inclusion of the most relevant 
mesoscopic fluctuations is necessary not only for the quantitative, but also for the qualitative description of systems with 
mesoscopic inhomogeneities. A tractable theory that would accurately predict properties of  systems with  
inhomogeneities on the mesoscopic length scale is still missing, however. 
A theory correctly  incorporating mesoscopic fluctuations should predict at least
a correct topology of the phase diagram and correct
properties of the disordered phase. 

Effects of periodic fluctuations of the order parameter (OP) $\phi$ are described by the Brazovskii
field theory~\cite{brazovskii:75:0}.
 However, the Brazovskii functional  $L_B[\phi]$
depends on free parameters,
and quantities such as compressibility
for a particular volume fraction $\zeta$ and $T$ cannot be determined,
unless a relation between the free parameters and measurable quantities is known. 
An attempt to express the free parameters in  $L_B[\phi]$ in terms of $T$ and $\zeta$ was made in 
Ref.\cite{ciach:08:1,ciach:13:0,ciach:11:0}. 
The theory developed in Ref.\cite{ciach:08:1,ciach:13:0,ciach:11:0} is  rather complex, however, 
and additional approximations are necessary to solve the obtained equations. 

The purpose of this work is  further development of
 the mesoscopic description of inhomogeneous systems that combines the DFT and field-theoretic 
approaches \cite{ciach:08:1,ciach:12:0,ciach:11:0}. We introduce a tractable approximation for the disordered phase. 
In sec.2 we
present a derivation of a  self-consistent equation for the correlation function, and the equation 
for the average volume fraction in terms of $T$ and $\mu$. The equations are obtained 
by the DFT methods, and
the contributions associated with the 
mesoscopic fluctuations can be calculated by the methods of field-theory. We show that  
the  term  $\propto\phi^3$, usually neglected in the Brazovskii-type theories,  
plays a very important role in fluids and soft-matter systems. 
In sec. 3 we derive the equations in the self-consistent Gaussian approximation. In
sec. 4 we limit ourselves to the disordered phase, make further assumptions  and obtain  much simpler equations
that can be solved easily.
In sec. 5  our equations are solved for a 1d model with the SALR potential. 
The results of our theory  are compared with the exact solutions obtained in Ref.\cite{pekalski:13:0} for a 1d lattice model.
We discuss the accuracy of our approximations in sec. 6.

\section{Derivation of the mesoscopic density functional theory}

We consider systems with competing interactions, where mesoscopic inhomogeneities occur on a length scale $\lambda$
significantly larger than the size of  molecules $\sigma$. Following ref.\cite{ciach:08:1,ciach:11:0}
we collect all the microscopic
states into disjoint sets; each set represents one mesostate described by a smooth function
$\zeta({\bf r})$. $\zeta({\bf r})$ is equal to
the fraction of the volume of a sphere with a center at ${\bf r}$ and a radius $\lambda\gg R\gg \sigma$ 
that is covered by the particles
 in each  microstate belonging to the set represented by $\zeta({\bf r})$. 
Another words, the mesoscopic state represents all the microscopic states that are obtained by changing 
the positions of the particles without changing the volume occupied by the particles in each mesoscopic region.
By fixing the mesostate
$\zeta({\bf r})$ we impose a constraint on the available microstates. The grand thermodynamic potential 
in the presence of this constraint
is  denoted by $\Omega_{co}[\zeta({\bf r})]$. $\exp(-\beta\Omega_{co}[\zeta({\bf r})])$ 
is equal to the sum of the Boltzmann factor $e^{-\beta H}$ 
over all microscopic states with frozen mesoscopic fluctuations.
In the above $\beta=1/(k_BT)$, $k_B$, $T$ and 
$H$ are the Boltzmann factor, temperature and the Hamiltonian respectively.
The grand potential in the presence of mesoscopic fluctuations is given by
\begin{equation}
 \Omega=-k_BT \ln \Xi
 \label{Omega}
\end{equation}
where
\begin{equation}
\label{Xi}
 \Xi=\int D\zeta e^{-\beta \Omega_{co}[\zeta]}
\end{equation}
and the functional integral is over all mesoscopic states that by definition are $\zeta<1$.
Since the summation of $e^{-\beta H}$  over all microstates compatible  with  $\zeta({\bf r})$ 
is included in $ e^{-\beta \Omega_{co}[\zeta]}$,
and in (\ref{Xi}) we perform the summation over all mesostates $\zeta({\bf r})$, $\Omega$ contains contributions from both, 
microscale and mesoscale fluctuations.

We introduce mesoscopic fluctuation by
\begin{equation}
 \phi:=\zeta-\bar\zeta
\end{equation}
where $\bar\zeta$ denotes the average volume fraction,
and rewrite (\ref{Omega}) in the equivalent form
\begin{equation}
\label{Om}
 \Omega=\Omega_{co}[\bar\zeta]-k_BT\ln\Bigg(\int D\phi e^{-\beta H_f}\Bigg)
\end{equation}
with
\begin{equation}
\label{Hf}
 H_f[\bar\zeta,\phi]=\Omega_{co}[\bar\zeta+\phi]-\Omega_{co}[\bar\zeta].
\end{equation}
Eq. (\ref{Om}) contains two contributions, and each of them is associated with the fluctuations on a different length scale.
In the first term the mesoscopic volume fraction has its equilibrium form. This term contains contribution from the 
fluctuations on the microscopic length scale. The second term contains the contributions from the fluctuations on the mesoscopic 
length scale, i.e. from different mesoscopic inhomogeneities that are thermally excited with the probability $ e^{-\beta H_f}/\Xi$.
When $\bar\zeta$ is the average volume fraction, then it follows that $\langle\phi\rangle_t=0$, where
\begin{equation}
\label{Xav}
 \langle X \rangle_f:=\frac{\int D \phi X e^{-\beta H_f}}{\int D \phi  e^{-\beta H_f}}.
\end{equation}
In the following  $\langle\phi\rangle_t=0$ is always assumed. 

 Our purpose here is the development of an approximate theory that allows for a determination of $\bar\zeta$ 
 for given temperature and chemical potential with the fluctuation contribution in (\ref{Om}) taken into account.
 We shall also find approximations for the 
 equation of state (EOS), the pair correlation
function   and the boundary of stability of the disordered phase.
Note that 
both terms on the RHS of Eq. (\ref{Om}) are functionals of $\bar\zeta$. The equilibrium volume fraction $\bar\zeta$
corresponds to the minimum of the grand potential
 i.e.  in equilibrium
 the first functional derivative of $\Omega[\bar\zeta]$ w.r.t. $\bar\zeta$
must vanish, and the second functional derivative must be positive definite.
We introduce the functional derivatives
\begin{eqnarray}
\label{Cn}
 C_n({\bf r}_1,...,{\bf r}_n):=\frac{\delta^n \beta\Omega}{\delta\bar\zeta({\bf r}_1)...\delta\bar\zeta({\bf r}_n)}=
 \\
 \nonumber
 C_n^{(0)}({\bf r}_1,...,{\bf r}_n)-\frac{\delta^n}{\delta\bar\zeta({\bf r}_1)...\delta\bar\zeta({\bf r}_n)}\ln\Big(\int D\phi e^{-\beta H_f[\bar\zeta,\phi]}
 \Big)
\end{eqnarray}
where (\ref{Om}) was used,
\begin{equation}
 C_n^{(0)}({\bf r}_1,...,{\bf r}_n)=\frac{\delta^n \beta\Omega_{co}}{\delta\bar\zeta({\bf r}_1)...\delta\bar\zeta({\bf r}_n)},
\end{equation}
 and we do not indicate the functional dependence of $C_n$ and $C_n^{(0)}$ on $\bar\zeta$.
 
  For $n=1,2$ we obtain from (\ref{Cn})
 \begin{equation}
 \label{C1}
C_1({\bf r})=C_1^{(0)}({\bf r}) +
 \langle \frac{\delta \beta H_f}{\delta \bar\zeta({\bf r})}\rangle_f,
\end{equation}
and
\begin{equation}
\label{C2}
 C_2({\bf r}_1,{\bf r}_2)=C_2^{(0)}({\bf r}_1,{\bf r}_2) +
 \langle \frac{\delta^2 \beta H_f}{\delta \bar\zeta({\bf r}_1)\delta \bar\zeta({\bf r}_2)}\rangle_f 
 - \langle\frac{\delta \beta H_f}{\delta \bar\zeta({\bf r}_1)}\frac{\delta \beta H_f}{\delta \bar\zeta({\bf r}_2)}\rangle_f^{con}
\end{equation}
where 
\begin{equation}
 \langle X Y \rangle_f^{con}:=\langle X Y \rangle_f -\langle X  \rangle_f\langle  Y \rangle_f.
\end{equation}
We neglect the fluctuation contribution to $C_n$ for $n\ge 3$, i.e.  we make the approximation
\begin{equation}
\label{Cn3}
 C_n=C_n^{(0)}  \hskip1cm {\rm for}  \hskip1cm n\ge 3.
\end{equation}

From (\ref{C1}) and the requirement $C_1({\bf r})=0$ we can obtain $\bar\zeta$ for given $T$ and the chemical potential $\mu$,
if we know the form of $\Omega_{co}$ and we can perform the functional
integrals in the second term on the RHS of (\ref{C1}). 

 We assume
the standard  local mean-field approximation for the grand potential with suppressed mesoscopic fluctuations,
\begin{equation}
\label{Omco}
 \Omega_{co}[\zeta]=U[\zeta] -TS[\zeta]-\mu N[\zeta].
\end{equation}
The entropy $S$ in the local density approximation for fixed mesoscopic volume fraction is given by 
\begin{equation}
-TS[\zeta]=\int d{\bf r} f_h(\zeta({\bf r})).
\end{equation}
Different approximations for the free-energy density of the reference (hard sphere) system, $f_h(\zeta)$, can be chosen. 
The internal energy  for fixed mesoscopic volume fraction is given by
\begin{equation}
\label{U}
 U[\zeta]=\frac{1}{2}\int d{\bf r}_1\int d{\bf r}_2\zeta({\bf r}_1)V({\bf r}_1-{\bf r}_2)\zeta({\bf r}_2),
\end{equation}
where  $V({\bf r}_1-{\bf r}_2)=u({\bf r}_1-{\bf r}_2)g({\bf r}_1-{\bf r}_2)$, with $u$ and $g$ denoting
the interaction potential and the pair distribution function respectively.  Note that 
we use volume fraction rather than density in (\ref{U}), therefore  we should re-scale 
the interaction potential $u({\bf r}_1-{\bf r}_2)$ by the factor $(6/\pi)^2$ to obtain the same energy
as in the standard theory. 
We also re-scale the chemical potential, $\bar\mu=(6/\pi)\mu$, so that 
\begin{equation}
\label{muN}
 \mu N[\zeta]=\bar\mu\int d{\bf r}\zeta({\bf r}),
\end{equation}
where $N[\zeta]$ is the number of particles for given $\zeta$.
In this local density approximation we have 
\begin{equation}
\label{C01}
  C_1^{(0)}({\bf r})= \int d{\bf r}_1 \zeta({\bf r}_1)\beta V({\bf r}_1-{\bf r})+A_1(\zeta({\bf r}))-\beta \bar\mu
\end{equation}
\begin{equation}
\label{C02}
 C_2^{(0)}({\bf r}_1,{\bf r}_2)=\beta V({\bf r}_1-{\bf r}_2)+A_2(\zeta({\bf r}_1))\delta({\bf r}_1-{\bf r}_2)
\end{equation}
and for $n\ge 3$
\begin{equation}
\label{C0n}
 C_n^{(0)}({\bf r}_1,...,{\bf r}_n)=A_n(\zeta({\bf r}_1))\delta({\bf r}_1-{\bf r}_2)...\delta({\bf r}_{n-1}-{\bf r}_n),
\end{equation}
where
\begin{equation}
\label{An}
 A_n(\zeta)=\frac{d^n\beta f_h(\zeta)}{d \zeta^n}.
\end{equation}

In order to perform the functional integrals in (\ref{Om}), (\ref{C1}) and (\ref{C2}) 
we expand $H_f[\bar\zeta,\phi]$ in a functional Taylor series w.r.t. $\phi$,
\begin{equation}
\label{Taylor2}
 \beta H_f[\bar\zeta,\phi]=  \beta H_0+\beta \Delta H
\end{equation}
with 
\begin{equation}
\label{Taylor20}
 \beta H_0[\bar\zeta,\phi]=  \frac{1}{2}
  \int d{\bf r}_1 \int d{\bf r}_2 \phi({\bf r}_1)C_2^{(0)}({\bf r}_1,{\bf r}_2)\phi({\bf r}_2)
\end{equation}
and
\begin{equation}
\label{Taylor2int}
 \beta \Delta H[\bar\zeta,\phi]=  \int d{\bf r} C_1^{(0)}({\bf r})\phi({\bf r})
 +\sum_{n\ge 3}\int d{\bf r} \frac{A_n(\zeta({\bf r}))}{n!} \phi({\bf r})^n.
\end{equation}
We truncate the expansion in (\ref{Taylor2int}) at the fourth order term. This truncation is justified, because by definition
the volume fraction is less than 1, and for
small fluctuations, $\phi\ll 1$, the higher order terms in (\ref{Taylor2int}) are irrelevant.  
The functional integral  in (\ref{Om})
can be extended to arbitrarily large functions $\phi$, because the large fluctuations are strongly damped in (\ref{Om}) 
by the Boltzmann factor $e^{-\beta H_f}$ for $H_f$ given in (\ref{Taylor2})-(\ref{Taylor2int}), and do not influence the results 
in a significant way. 

Note that when $H_f$ is given by (\ref{Taylor2})-(\ref{Taylor2int}), the  fluctuation contributions in (\ref{C1})  and (\ref{C2}) 
can be expressed in terms of $\langle\phi^n({\bf r})\rangle$ and  $\langle\phi^n({\bf r}_1)\phi^m({\bf r}_2)\rangle$.
Thus, it is necessary to calculate the correlation functions for the mesoscopic fluctuations of the volume fraction in order to 
obtain the expressions for $C_1$ and $C_2$. In order to calculate these functions field-theoretic methods could be used. 
In the next section  we present an approximation that allows to determine $C_2$ and $\langle\phi({\bf r}_1)\phi({\bf r}_2)\rangle$
in a relatively simple way.

\section{Self-consistent Gaussian approximation}

In practice
we can calculate  the Gaussian functional integrals only.  
We approximate $H_f$
by an effective functional that is quadratic in the fluctuation $\phi$,
\begin{eqnarray}
\label{HG}
 \beta H_f\approx\beta  H_G= \frac{1}{2}\int d{\bf r}_1\int d{\bf r}_2 \phi({\bf r}_1)C({\bf r}_1,{\bf r}_2)\phi({\bf r}_2).
\end{eqnarray}
In this approximation the form of $C$ differs from $C_2^{(0)}$, so that  the effect
of  $\Delta H$ (see (\ref{Taylor2})-(\ref{Taylor2int}))
is indirectly included
in the average quantities (see (\ref{Xav})). Note that when $H_f$ is approximated by $H_G$, 
then  the correlation function $G({\bf r}_1,{\bf r}_2):=\langle\phi({\bf r}_1)\phi({\bf r}_2)\rangle$ is given by
\begin{equation}
\label{CG}
 \int d {\bf r}_2 G({\bf r}_1,{\bf r}_2)C({\bf r}_2,{\bf r}_3)=\delta({\bf r}_1,{\bf r}_3) .
\end{equation}
 The best approximation for
$C({\bf r}_1,{\bf r}_2)$ is such that $G$ in (\ref{CG}) is as close as possible to the {\it exact} correlation function. 
This means that $C$ should be as close as possible to the {\it exact}  inverse correlation function~\cite{evans:79:0}. 
Here, we assume that  $C= C_2$. 
Note that in the Gaussian approximation the $2n$-point correlation functions
can be expressed in terms of products of the two-point correlation functions. Thus, because the fluctuation contribution in 
Eq.(\ref{C2}) consists of terms proportional to $\langle\phi^n({\bf r}_1)\phi^m({\bf r}_2)\rangle$, it relates $C_2=C$ with $G$, and
Eq.(\ref{CG}) relates $G$ with $C$. 
In this self-consistent Gaussian
approximation  
$\langle X\rangle_f$ in Eq.(\ref{C2}) is replaced by 
$\langle X\rangle=\int D\phi X \exp(-\beta H_G)/ \int D\phi  \exp(-\beta H_G)$, with $H_G$ given in (\ref{HG}).

In order to calculate the functional integrals in Eqs.(\ref{C1}) and (\ref{C2}) in the above Gaussian approximation,
we note that from (\ref{HG}), (\ref{Cn}),
(\ref{Cn3}) and  (\ref{C0n}) it follows that
\begin{equation}
 \frac{\delta\beta H_G}{\delta\zeta({\bf r})}=\frac{1}{2} A_3(\zeta({\bf r}))\phi ({\bf r})^2
\end{equation}
and
\begin{equation}
 \frac{\delta^2\beta H_G}{\delta\zeta({\bf r}_1)\delta\zeta({\bf r}_2)}=
 \frac{1}{2} A_4(\zeta({\bf r}_1))\phi ({\bf r}_1)^2\delta ({\bf r}_1-{\bf r}_2).
\end{equation}
After inserting the above equations in (\ref{C1}) and (\ref{C2}), using  (\ref{C01}) - (\ref{An})
and performing the Gaussian integrals, we obtain
the main results of the self-consistent Gaussian approximation
\begin{equation}
\label{mu}
\beta \bar\mu=\int d{\bf r}_1 \zeta({\bf r}_1)\beta V({\bf r}_1-{\bf r})+
A_1(\zeta({\bf r}))+\frac{A_3(\zeta({\bf r}))}{2}G({\bf r},{\bf r})
\end{equation}
and
\begin{equation}
\label{selfcon}
 C({\bf r}_1,{\bf r}_2)=C^{(0)}_2({\bf r}_1,{\bf r}_2)+\frac{A_4(\zeta({\bf r}_1))}{2}G({\bf r}_1,{\bf r}_2)\delta({\bf r}_1-{\bf r}_2)
 -\frac{A_3(\zeta({\bf r}_1))A_3(\zeta({\bf r}_2))}{2}G^2({\bf r}_1,{\bf r}_2).
\end{equation}
Eqs. (\ref{selfcon}) and (\ref{CG}) have to be solved self-consistently. 
The last term in (\ref{mu})  and the  last two terms in (\ref{selfcon})
represent the contributions from the mesoscopic fluctuations.  
We should stress that because of the coarse graining, the correlation function for the mesoscopic volume fraction, 
$G({\bf r}_1,{\bf r}_2)$, differs from 
the correlation function for the microscopic density.
In Ref.\cite{ciach:08:1,ciach:11:0}
it was shown that $G({\bf r}_1,{\bf r}_2)$ 
represents the microscopic correlations  averaged  over  mesoscopic regions (smaller than the scale of inhomogeneities) 
around the points ${\bf r}_1$ and ${\bf r}_2$.
As a consequence, $G({\bf r}_1,{\bf r}_2)\ne 0$ for $ {\bf r}_1={\bf r}_2$, because it contains
contributions from the microscopic correlations
between the particles with the centers belonging to the sphere of the radius $R$ larger than $\sigma/2$, i.e. between the particles 
which do not overlap.
Note also that
the fluctuation contribution  leads to a shift
of the average volume fraction for given $T$ and $\bar\mu$ compared  to the MF prediction.

\begin{figure}
\vskip0.5cm
 \includegraphics[scale=0.3]{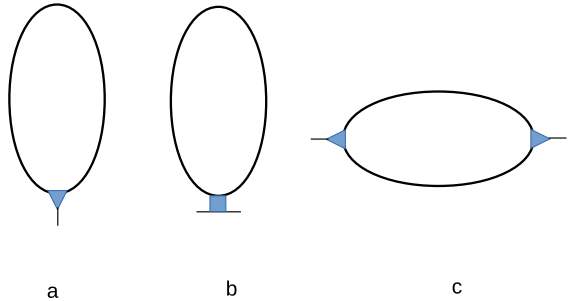}
 \label{Feynman}
 \caption{The Feynman diagrams representing the fluctuation contribution in  Eq. (\ref{mu}) (a) and in Eq.(\ref{selfcon}) (b) and (c).
 Thick line connecting vertices at ${\bf r}_1$ and ${\bf r}_2$  represents $G({\bf r}_1,{\bf r}_2)$. 
 The vertex with a triangle or the square  represents $A_3$ or $A_4$ respectively. }
 \vskip0.5cm
\end{figure}

Our result (\ref{selfcon}) agrees with the well known expression 
in the field theory 
\cite{brazovskii:75:0} at the self-consistent one-loop approximation.
The corresponding Feynman diagrams are shown in Fig.1.
In the standard field-theoretic approaches, however, $A_n$ are free model parameters, and 
the last term  in (\ref{selfcon}) is usually 
neglected. In our theory $A_3$ is a function of $\zeta$ and vanishes only at a single value of the volume fraction, corresponding
to the critical point in a  system with attractive interactions. 
In our theory $A_3$ plays a very important role for the relation between the average volume fraction and the chemical potential.

\section{Self-consistent Gaussian approximation for the disordered  phase}

In this section we restrict our considerations to a disordered isotropic phase with the mesoscopic volume fraction independent of the position,
and to  isotropic interactions, i.e.  $V({\bf r}_1-{\bf r}_2)$ depending only on  $r=|{\bf r}_1-{\bf r}_2|$. 
In this phase $G$ and $C$ depend only on  $r=|{\bf r}_1-{\bf r}_2|$, and we simplify the notation, introducing 
$G(r)\equiv G({\bf r}_1,{\bf r}_2)$ and $C(r)\equiv C({\bf r}_1,{\bf r}_2)$.
Due to the translational invariance,  Eq.(\ref{mu}) takes the form
\begin{equation}
\label{muhom}
\beta \bar\mu=\bar\zeta\int d{\bf r}\beta V(r)+
A_1(\bar\zeta)+\frac{A_3(\bar\zeta)}{2}{\cal G},
\end{equation}
and we can write Eq.(\ref{selfcon}) in  Fourier representation
\begin{equation}
\label{selfconF}
 \tilde C(k)= \tilde C^{(0)}_2(k)+\frac{A_4(\bar\zeta)}{2}{\cal G}-\frac{A_3^2(\bar\zeta)}{2}\tilde D(k),
\end{equation}
where $\tilde F$ denotes the Fourier transform of $F$, and we have introduced 
\begin{equation}
 {\cal G}=G({\bf r},{\bf r})
\end{equation}
and
\begin{equation}
\label{Dk}
 \tilde D(k)=\int d{\bf r} G^2(r)e^{i{\bf k}\cdot{\bf r}}.
\end{equation}
 In the disordered isotropic phase  Eq.(\ref{CG}) 
takes the form 
\begin{equation}
\label{CGF}
 \tilde C(k)\tilde G(k)=1.
\end{equation}

 In the context of fluids with 
 competing interactions, $A_3\ne 0$ except from the critical value of the volume fraction, $\zeta_c$.
 $A_3(\zeta)<0$ for $\zeta<\zeta_c$, and $A_3(\zeta)>0$ for $\zeta>\zeta_c$. Thus, the fluctuation contribution in (\ref{muhom})
 leads to decreased and increased value of the chemical potential for  $\zeta<\zeta_c$ and $\zeta>\zeta_c$ respectively compared
 to the MF prediction, provided that ${\cal G}>0$. This effect of mesoscopic fluctuations 
 is independent of the details of the interaction potential. Note that the value of ${\cal G}$ is the larger the stronger are the 
 inhomogeneities on the mesoscopic length scale. 

In order to solve Eq.(\ref{selfconF}) for a particular system, we need to know the
form of the interaction potential. However, we can make general qualitative or semi-quantitative predictions for a class
of systems with interactions such that
 $\tilde V(k)$ has a pronounced minimum at $k=k_0>0$. The  wavenumbers of the
dominant mesoscopic fluctuations  correspond to the maximum of 
the Boltzmann factor $\exp(-\beta H_G)$, and we assume that only  $k\approx k_0$ are relevant when the peak 
in $\exp(-\beta\tilde V(k))$ is pronounced and 
narrow \cite{ciach:08:1,ciach:11:0}. 
We focus on such systems, and make the approximation
\begin{equation}
\label{tildeCa}
 \tilde C^{(0)}_2(k)\approx c_0 +v_0(k^2-k_0^2)^2
\end{equation}
where
\begin{equation}
\label{c0}
 c_0 = A_2+\beta\tilde V(k_0)
\end{equation}
and
\begin{equation}
\label{v0}
 v_0=\frac{\beta\tilde V''(k_0)}{8k_0^2}.
\end{equation}
We  have used (\ref{C02}) and have made the approximation $k+k_0\approx 2k_0$, valid in the neighborhood of $k_0$.
The above assumption is justified for an even function of $k$ in the neighborhood of $k_0>0$, where the first derivative w.r.t. $k$
is $\tilde V'(k_0)=0$, and the second derivative is $\tilde V''(k_0)>0$. 
The results obtained with this assumption concern any system with inhomogeneities on the
mesoscopic length scale $\sim 2\pi/k_0$ strongly  favoured energetically compared to inhomogeneities on different length scales. 
Thus, the results presented below concern systems with a deep minimum of $\tilde V(k)$.
For potentials with a shallow minimum the approximation (\ref{tildeCa}) can be an oversimplification.

The equations (\ref{selfconF}) - (\ref{CGF}) are  still rather difficult. It is thus reasonable to make further approximations
in order to verify if the predictions of this theory are correct at least on a qualitative level. Instead of solving 
numerically Eqs.(\ref{selfconF}) - (\ref{CGF}), we  postulate a particular form of $\tilde C(k)$, and determine 
the parameters in the postulated expression.
Experimental results for the structure factor 
in systems with mesoscopic inhomogeneities are reasonably well described by a function inversely proportional 
to $\tilde C^{(0)}_2(k)$ given in Eq.(\ref{tildeCa}),
with $c_0, k_0$ and $v_0$ treated as fitting parameters. 
Based on this observation we assume for $\tilde C(k)$ the same  form as in Eq.(\ref{tildeCa}), but with renormalized
parameters, i.e. we postulate
\begin{equation}
\label{tildeCaren}
 \tilde C(k)\approx c_r +v_r(k^2-k_r^2)^2.
\end{equation}

In order to determine the renormalized parameters we need 3 equations for the 3 unknowns.
 We require that  the first derivative of the function given by Eq.(\ref{selfconF}) vanishes at $k=k_r$.
 The value of this function at its minimum  is $c_r$. Finally, we require that
 the second derivatives  at $k_r$ of both expressions, Eqs.(\ref{selfconF})
 and (\ref{tildeCaren}),  are the same. 
 The three requirements guarantee that the shape of $\tilde C(k)$, Eq.(\ref{selfconF}),  is reproduced by
 Eq.(\ref{tildeCaren}) at least near the minimum.
 From these requirements we obtain  the set of equations
 \begin{equation}
 \label{cr}
  c_r=c_0 + v_0(k_r^2-k_0^2)^2 +\frac{A_4}{2} {\cal G} -\frac{A_3^2}{2} \tilde D(k_r)
 \end{equation}
\begin{equation} 
\label{kr}
 4v_0(k_r^2-k_0^2)k_r=\frac{A_3^2}{2} \tilde D^{'}(k_r)
\end{equation}
\begin{equation}
 \label{vr}
 8v_rk_r^2=4v_0(3k_r^2-k_0^2)-\frac{A_3^2}{2} \tilde D^{''}(k_r).
\end{equation}

Note that in the standard Brazovskii approximation, i.e. with $A_3=0$,
from (\ref{kr}) and (\ref{vr}) we obtain $k_r=k_0$, $v_r=v_0$, and Eq.(\ref{cr}) can be easily solved analytically \cite{ciach:12:0}.

The pressure $p=- \Omega /{\cal V}$, where ${\cal V}$ is the system volume, 
in this theory is given by (see (\ref{Omega}), (\ref{Omco})-(\ref{muN}))
\begin{eqnarray}
\label{p}
 p=-\frac{1}{2}\tilde V(0)\bar \zeta^2 - f_h(\bar\zeta) +\bar\mu \bar\zeta +\frac{k_BT}{{\cal V}} \ln\Bigg(
\int D\phi e^{-\beta H_G} \Bigg).
\end{eqnarray}
We  express $\bar \mu$  in terms of $\bar\zeta$ according to Eq.(\ref{muhom}), 
evaluate the functional integral and obtain from (\ref{p}) the EOS of the form
\begin{eqnarray}
 \label{EOS}
  p=\frac{1}{2}\tilde V(0)\bar \zeta^2 - f_h(\bar\zeta)+k_BT A_1(\bar\zeta)\bar\zeta +k_BT\Bigg[
  \frac{A_3(\bar\zeta)\bar\zeta{\cal G}}{2} -
\int \frac{d {\bf k}}{2(2\pi)^d} \ln\Big(\frac{\tilde C(k)}{2\pi}\Big)
  \Bigg].
\end{eqnarray}

The integral in (\ref{EOS}) is over the spectrum of mesoscopic fluctuations, and is cutoff-dependent for $\tilde C(k)$ 
given in Eq.(\ref{tildeCaren}). 
In the mesoscopic theory the cutoff is naturally provided by the scale $R$ of coarse-graining. However, the 
pressure should be independent of the coarse-graining procedure and the cutoff-dependent part should cancel against the 
neglected cutoff-dependent contribution
to $\Omega_{co}$. We
find that the cutoff-independent contribution to the integral in (\ref{EOS})  is proportional to $  \sqrt{c_r/v_r}$
for  the 3d and 1d systems. Note that the approximations (\ref{tildeCa})
and (\ref{tildeCaren}) are valid for strong inhomogeneities 
at a well-defined length scale, i.e. for 
 $c_r/v_r\ll 1$. Since ${\cal G}\propto 1/\sqrt{c_r}$~\cite{brazovskii:75:0} (see also the next section),
in the range of validity
of the approximations, the last term in (\ref{EOS}) is negligible compared to the fluctuation contribution proportional 
to  ${\cal G}$, and will be disregarded.
 
\section{Results for the SALR potential in 1d}
In this section we  test the accuracy of the self-consistent Gaussian approximation with the further approximations
described in sec. 4 by comparing our results 
with the exact solutions obtained in Ref.~\cite{pekalski:13:0} for a 1d lattice model with competing attractive and repulsive interactions
between the first and the third neighbors respectively. 
 In the lattice model solved exactly in 1d the ratio between the third-neighbor repulsion $J_2$ 
 and the first-neighbor attraction $J_1$ 
 is denoted by $J$,
 and the interaction potential in Fourier representation takes the form
 \begin{equation}
 \label{Vlatt}
  \beta
  \tilde V(k)= 2\beta^*\Big( J \cos(3k)-\cos(k)\Big)
 \end{equation}
where $\beta^*=1/T^*$, and $T^*=k_BT/J_1$ is  temperature in units of the   first-neighbor attraction.
 $\tilde V(k)$ in Eq.(\ref{Vlatt}) assumes a negative minimum for $k_0>0$ if $J>1/9$.  By $\tilde V^* (k)$, $\mu^*$  and $p^*$
 we shall denote the corresponding quantity in units of $J_1$.

 From the form of $\tilde C^{(0)}_2$ (Eqs.(\ref{tildeCa})-(\ref{v0}))
 it immediately follows that in MF the disordered phase is unstable
for $T^*<-\tilde V^*(k_0)/A_2(\zeta)$. 
The explicit form of $A_2(\zeta)$ follows from  (\ref{An}) and 
the reference-system  free energy of the lattice-gas form 
(note that in 1d the volume fraction and the dimensionless number density are identical),
\begin{equation}
\label{fh}
  \beta f_h(\zeta)=\zeta \ln \zeta + (1-\zeta)\ln(1-\zeta).
\end{equation}
This well known MF result \cite{archer:08:0,ciach:08:1,ciach:10:1} 
is incorrect, since in 1d models with short-range interactions there are no phase transitions for $T>0$ \cite{landau:89:0}.
Exact results~\cite{pekalski:13:0}  indicate, however
that the disordered phase is strongly inhomogeneous in the phase-space region where MF predicts
its instability.
Thus, it is interesting to verify predictions of our theory for $T^*<-\tilde V^*(k_0)/A_2(\zeta)$.
 We shall focus on  $T^*<-\tilde V^*(k_0)/A_2(\zeta)$ for $J=3$ and $J=1/4$, 
 because most of the exact results of Ref.~\cite{pekalski:13:0} concern $J=3$ and $J=1/4$.
 The phase-space region of interest is enclosed by the solid and dashed lines in Fig.2 respectively. 
 
 The two chosen values of $J$
 correspond to systems with qualitatively different properties. In the 
 case of $J=1/4$ only the empty or the fully occupied lattice is present 
 in the ground state (GS) for small or large values of $\mu^*$ respectively. In a system with the short-range 
attraction stronger than the long-range repulsion,
exact results  \cite{pekalski:13:0} 
show stability of the disordered phase with oscillatory and monotonic decay of correlations at large 
and at very small $T^*$ respectively. The compressibility at $\zeta\approx 1/2$ is very large at low $T^*$, 
signaling the approach to the phase separation at $T^*=0$.
 
 For strong repulsion ($J>1$)
clusters composed of 3 particles separated by 3 empty sites are  favoured energetically,
since there are as many attractions (occupied nearest-neighbour sites) as possible 
in the absence of repulsion. Such a periodic phase is stable for intermediate values of  $\mu^*$ in the GS.
Exact results \cite{pekalski:13:0} 
show that for $J=3$ the correlation function exhibits oscillatory decay 
with the  wavelength 
$\sim 6$.
The correlation length is several 
orders of magnitude larger
than the particle size for $\zeta\approx 1/2$ and  $T^*\ll -\tilde V(k_0)/A_2(\zeta)$.
Both the correlation length and the amplitude of $G(r)$ decay to much smaller values when 
the line of MF instability is approached. The compressibility at low $T^*$ is very small for $\zeta\approx 1/2$ and very large 
otherwise. The small compressibility signals formation of regularly
distributed clusters that repel each other upon compression of the system (when their
separation becomes shorter than the range of repulsion). The large compressibility signal the approach to the phase transitions 
between the periodic phase and 
the dilute gas or dense liquid phases at $T=0$. 

Our purpose is the verification if the mesoscopic theory in the self-consistent
Gaussian approximation and with the further assumptions made in sec.4 is able to reproduce the above exact results. 
  In order to calculate the correlation function and the $\mu^*(\zeta)$ and $p^*(\zeta)$ lines, we have to 
  solve Eqs.(\ref{cr})-(\ref{vr}). To solve these equations  we need the form of $\tilde D(k)$, 
  which depends on $G(r)$ (see (\ref{Dk})).
  In real space representation the correlation function $\tilde G^{(0)}(k)=1/\tilde C^{(0)}_2(k)$ 
 for $\tilde C^{(0)}_2(k)$ approximated by Eq.(\ref{tildeCa})
 takes   the form
\begin{equation}
\label{G1d}
  G^{(0)}(r)=A_0 e^{-r/\xi_0}\Big(
  \alpha_0\cos(\alpha_0 r)+\xi_0^{-1}\sin(\alpha_0r)
  \Big),
\end{equation}
 where the correlation length is 
\begin{equation}
\label{xi0}
 \xi_0= 2\alpha_0\sqrt{\frac{v_0}{c_0} },
\end{equation}
 and  the wavenumber of the damped oscillations $ \alpha_0$ and the amplitude $A_0$  are given by 
\begin{equation}
 \alpha_0^2=\frac{k_0^2 + \sqrt{k_0^4 + c_0/v_0}}{2},
\end{equation}
 and 
\begin{equation}
 A_0^2=\frac{1}{4v_0c_0(k_0^4+c_0/v_0)}.
\end{equation}
 Eq.(\ref{G1d}) was obtained for  continuum space.
Our goal is to verify if our approximate theory reproduces the qualitative features of the exact results, therefore
we assume the same form for the lattice model because of its simplicity.
 According to the ansatz (\ref{tildeCaren}), $G(r)$ has the form given in Eq.(\ref{G1d}),  
but with the parameters
$A_0,\xi_0,\alpha_0$ replaced by the renormalized ones, $A_r,\xi_r,\alpha_r$ respectively. 
The relation of the parameters $A_r,\xi_r,\alpha_r$
with $c_r, k_r, v_r$ is analogous to the relation between $A_0,\xi_0,\alpha_0$ and  $c_0, k_0, v_0$ given above. 
The explicit form of $\tilde D(k)$  can be easily obtained, and reads
\begin{equation}
\label{D1}
 \tilde D(k)=\frac{A_r^2}{2\xi_r}\Bigg[
 \frac{\alpha_r^2-\xi_r^{-2} - 2\alpha_r(k- 2\alpha_r)}{(k- 2\alpha_r)^2 +4\xi_r^{-2}} +
 \frac{\alpha_r^2-\xi_r^{-2} + 2\alpha_r(k+ 2\alpha_r)}{(k+ 2\alpha_r)^2 +4\xi_r^{-2}} +
 \frac{4(\alpha_r^2+\xi_r^{-2})}{k^2 + 4 \xi_r^{-2}}.
 \Bigg]
\end{equation}
The above and Eqs. 
(\ref{An}), (\ref{fh}), (\ref{c0}) - (\ref{v0}) 
allow us
to solve Eqs. (\ref{cr})-(\ref{vr}) numerically.

 \begin{figure}
 \vskip0.5cm
 \includegraphics[scale=0.33]{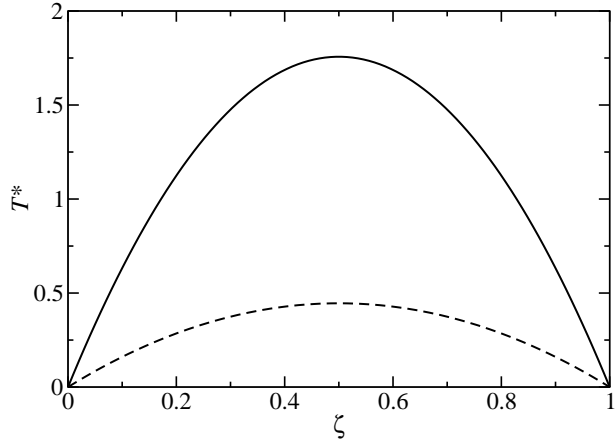}
 \caption{The lines of instability of the disordered phase in the 1d lattice model of Ref.\cite{pekalski:13:0} 
 in the MF approximation for the repulsion to attraction ratio $J=3$ (solid line) and $J=1/4$ (dashed line). 
 Temperature is in units of the nearest-neighbor attraction ($T^*=k_BT/J_1$).
 The volume fraction of particles $\zeta$ is dimensionless.
 We study properties of the disordered phase inside the region where this phase is unstable in MF, i.e. below the shown lines.}
 \vskip0.5cm
\end{figure}

Before presenting the results we first verify if the assumption that $\tilde C_2^{(0)}(k)$ 
can be approximated by Eq.(\ref{tildeCa}) is valid for this model. 
 We Taylor expand $\tilde V(k)$ given by 
Eq.(\ref{Vlatt}) in terms of $k^2$ about its minimum at $k^2=k_0^2$. The form of $\tilde V(k)$ (Eq.(\ref{Vlatt})) and
the Taylor expansion truncated as in Eq.(\ref{tildeCa})  
are shown in Fig.3 for $J=3$ and $J= 1/4$.

\begin{figure}
\vskip0.5cm
  \includegraphics[scale=0.33]{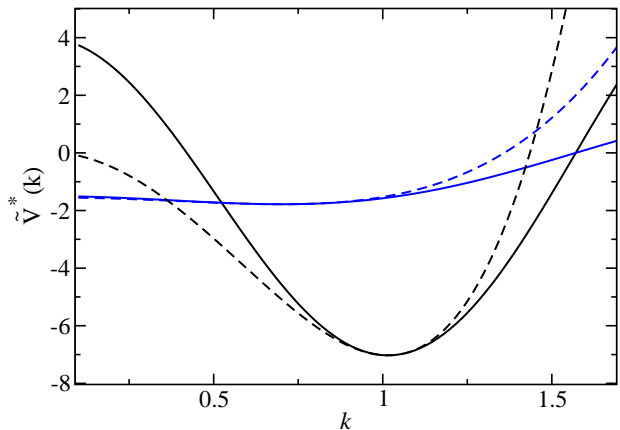}
  \caption{$\tilde V^*(k)$ (Eq.(\ref{Vlatt})) for $J=3$  and $J=1/4$ (solid lines) and the Taylor expansion leading to 
  the approximate form of $ \tilde C^{(0)}_2$ (\ref{tildeCa})
  (dashed lines). The deeper minimum corresponds to
   $J=3$. $\tilde V^*(k)$ is in units of the nearest-neighbor
   attraction $J_1$ and $k$ is in units of $1/\sigma$, with $\sigma$ the particle diameter.
  }
  \vskip0.5cm
\end{figure}

We can see that our approximation is good for $k\approx k_0$. However, 
the approximate formula overestimates and underestimates the effects of fluctuations with 
the wavelengths $k<k_0$ and $k>k_0$ respectively. Contrary to the assumption of a deep minimum (sec.4), 
in the case of $J=1/4$ the minimum of $\tilde V(k)$ is very shallow.

In the second step we verify the  ansatz (\ref{tildeCaren}). In our self-consistent Gaussian approximation 
the inverse correlation function $\tilde C(k)$ should satisfy both, Eq.(\ref{tildeCaren}) and Eq.(\ref{selfconF}). We
compare  Eqs. (\ref{tildeCaren}) and (\ref{selfconF}) 
with each other, and 
  with the MF version (\ref{tildeCa}). In
Fig.4a we show  Eqs. (\ref{tildeCaren}), (\ref{selfconF}) and (\ref{tildeCa})
  for $\bar\zeta=0.24$, $T^*=0.1$ and $J=3$, and in Fig.4b for $\bar\zeta=0.41$, $T^*=0.1$ and $J=1/4$.
Since $-\tilde D(k)$ has minima at $k=0,2\alpha_r$ (see (\ref{D1})) and in the ansatz (\ref{tildeCaren}) 
there is a single minimum at $k=k_0$,
the agreement between Eqs.(\ref{tildeCaren}) and (\ref{selfconF}) becomes worse for increasing  $|A_3|$, i.e. 
for increasing $|\bar\zeta -1/2|$. Hence, the ansatz can be acceptable only for a limited range of $\zeta$.

For $J=1/4$ it turns out that solutions of (\ref{cr})-(\ref{vr}) exist only in the central region of
the MF instability of the disordered phase. For  $T^*\approx 0.1$ the solutions  exist
only for $0.4<\zeta< 0.6$. As shown in Fig.4b, however, the agreement between (\ref{tildeCaren}) and (\ref{selfconF}) is 
still not satisfactory for $k<k_r$, except in the vicinity of $\zeta=1/2$.
Thus, the ansatz (\ref{tildeCaren})
is incorrect and either a different form of $\tilde C(k)$ should be assumed, or numerical solution of (\ref{selfconF})
is necessary.

For $J=3$ the agreement between the ansatz (\ref{tildeCaren}) and the Eq.(\ref{selfconF}) is fair for a broader range of $k$. 
We shall consider $0.1<\bar\zeta<0.9$,
since our approximation  (\ref{tildeCaren}) 
becomes increasingly oversimplified when  small and large volume fractions are approached.
We have verified that  the set of equations 
(\ref{cr})-(\ref{vr}) has solutions with $c_r>0$ 
below the line of MF instability  shown in Fig.2. This indicates the lack of instability 
of the disordered phase, in agreement with exact results (see (\ref{tildeCaren})).

\begin{figure}
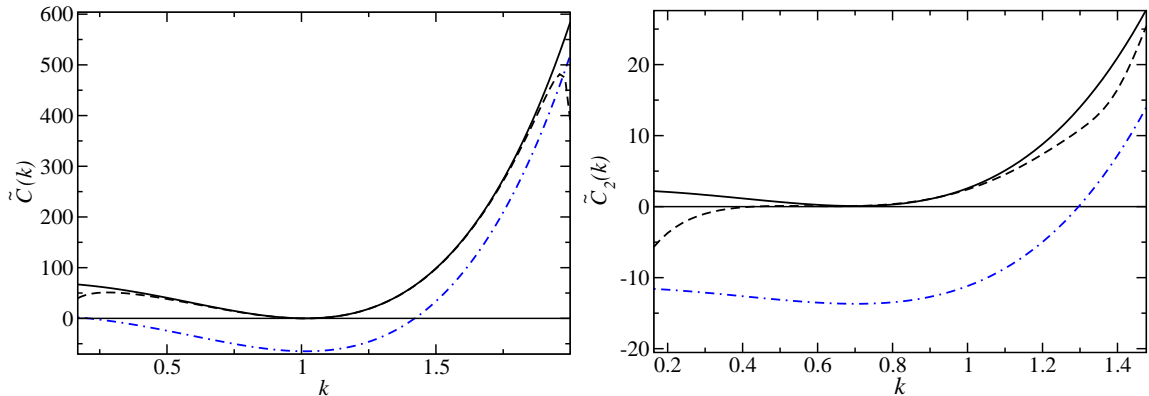

\vskip1cm
  \includegraphics[scale=0.3]{fig4a.eps}
   \includegraphics[scale=0.3]{fig4b.eps}
  \caption{Comparison of the inverse correlation function $C$ 
  in MF (dash-dotted line), and as given by Eqs.(\ref{selfconF}) and (\ref{tildeCaren}),
  (dashed and solid line respectively). Top  panel: $J=3$, $\bar\zeta=0.24$ and $T^*=0.1$. Bottom panel: $J=1/4$, $\bar\zeta=0.41$ 
  and $T^*=0.1$. 
  The parameters in  (\ref{tildeCaren}) satisfy Eqs. (\ref{cr})-(\ref{vr}).
 Note the negative values in MF, indicating the instability of the disordered phase, and the very small positive values in our theory.
 Note also that in the case of $J=1/4$ 
  the assumption (\ref{tildeCaren}) is not satisfactory  for $k<k_r$.  $C$ is dimensionless and 
  $k$ is in units of $1/\sigma$, with $\sigma$ the particle diameter.}
  \vskip1cm
\end{figure}

Let us focus on the structure of the disordered phase. 
In Fig.5 we show the correlation length $\xi_r$ and the amplitude $A_r$ of
the correlation function as functions of  $\zeta$ for three different temperatures.
The correlation function (see Eq.(\ref{G1d}) with renormalized parameters) 
is shown  for $\zeta=0.24$ and two values of temperature, $T^*=0.1$ and $T^*=0.7$
in Fig.6. 
 For low $T^*$ the correlation length is very large for  $\zeta\approx 1/2$, and rapidly decreases when $|\zeta- 1/2|$ increases,
 in agreement with the results of Ref.\cite{pekalski:13:0}. 
On the quantitative level, however, the accuracy of the approximation decreases for decreasing $T^*$. For $T^*\sim 0.5$ we obtain
semi-quantitative agreement between the approximate and exact results, but for $T^*\sim 0.1$ 
the correlation length is significantly smaller and the amplitude is significantly larger than obtained in Ref.\cite{pekalski:13:0}. 
$A_r>1$ (Fig.6) is an artifact of our approximations. Note, however that in MF $A_0\to\infty$ for 
$T^*\to -\tilde V(k_0)/A_2(\zeta)$ from above, so the improvement in our theory is significant.

\begin{figure}
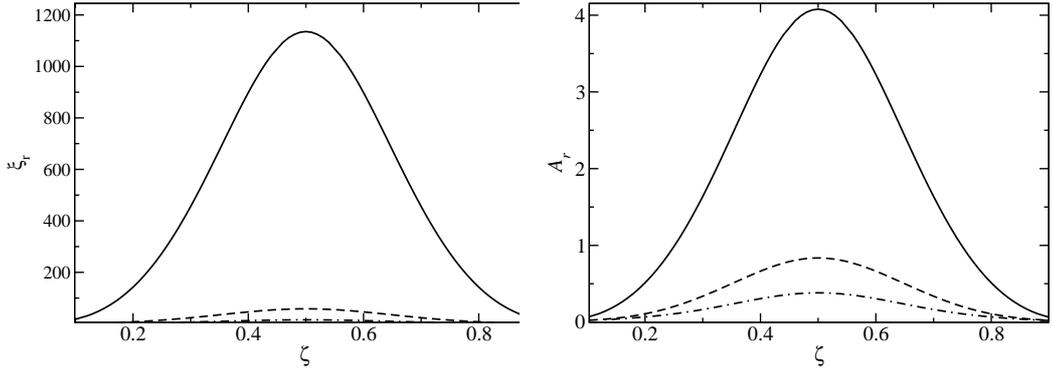

\vskip1cm
\label{xiam}
 \includegraphics[scale=0.28]{fig5a.eps}
 \includegraphics[scale=0.28]{fig5b.eps}
 \caption{The correlation length $\xi_r$ (left panel) and the amplitude $A_r$ (right panel) 
 as  functions of the volume fraction $\zeta$ (dimensionless) for $J=3$. From the top to the 
 bottom line $T^*=0.1,0.4,0.7$. 
 $\xi_r$ is in units 
  of $1/\sigma$, with $\sigma$ the particle diameter, and  $A_r$ is dimensionless.}
  \vskip1cm
\end{figure}

\begin{figure}
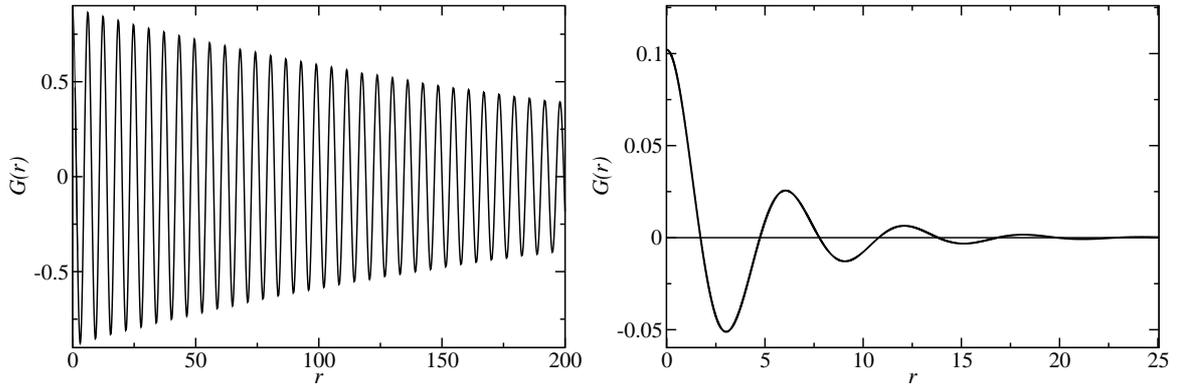

\vskip1cm
\label{corfun}
 \includegraphics[scale=0.3]{fig6a.eps}
 \includegraphics[scale=0.3]{fig6b.eps}
 \caption{Correlation function for $\zeta=0.24$ in the case of $J=3$. $T^*=0.1$ (left) and $T^*=0.7$ (right).
  $G(r)$ is dimensionless and $r$ is  in units of  the particle diameter $\sigma$. }
  \vskip1cm
\end{figure}

In Fig.7
the wavenumber $k_r$ at the minimum of $\tilde C(k)$ (maximum of the structure factor) is shown for $J=3$ and $T^*=0.4$.
It  decreases 
slightly for increasing $|\bar\zeta - 0.5|$, indicating increasing wavelength of inhomogeneities, in agreement with exact results~
\cite{pekalski:13:0}. However, the magnitude of $k_r$ obtained in Ref.\cite{pekalski:13:0} is slightly smaller. 

\begin{figure}
\vskip1cm
 \includegraphics[scale=0.33]{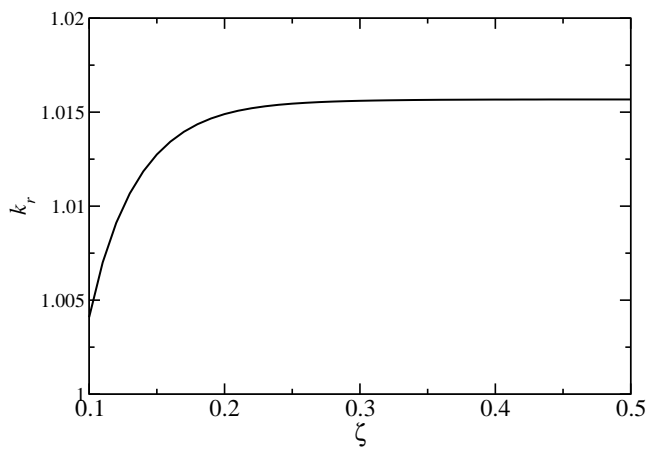}
 \caption{The wavenumber $k_r$  
 corresponding to the maximum of the structure factor (minimum of $\tilde C(k)$) as a function of
 the average volume fraction $\zeta$  (dimensionless) for $J=3$ and $T^*=0.4$.
 $k_r$ is in units of $1/\sigma$, with $\sigma$ the particle diameter.}
 \vskip1cm
\end{figure}

Let us focus on the effect of mesoscopic fluctuations on thermodynamic and mechanical properties in the 1d model.
The explicit expressions for the chemical potential and the EOS for $f_h$ given in Eq. (\ref{fh}) have the form (see (\ref{muhom}) 
and (\ref{EOS}))
\begin{eqnarray}
\label{mu1d}
  \mu^*=\tilde V^*(0) \zeta +T^* \ln\Bigg( \frac{\zeta}{1-\zeta}
 \Bigg) +T^* \frac{{\cal G}(2\zeta -1)}{2\zeta^2(1-\zeta)^2 }
\end{eqnarray}
and
\begin{eqnarray}
 \label{EOS1d}
  p^*=\frac{1}{2}\tilde V^*(0)\bar \zeta^2 -T^*\ln(1-\bar\zeta) +T^*
  \frac{{\cal G}(2\zeta -1)}{2\zeta(1-\zeta)^2} .
\end{eqnarray}
As discussed at the end of sec.4, we have  neglected the last term in Eq.(\ref{EOS}).

\begin{figure}
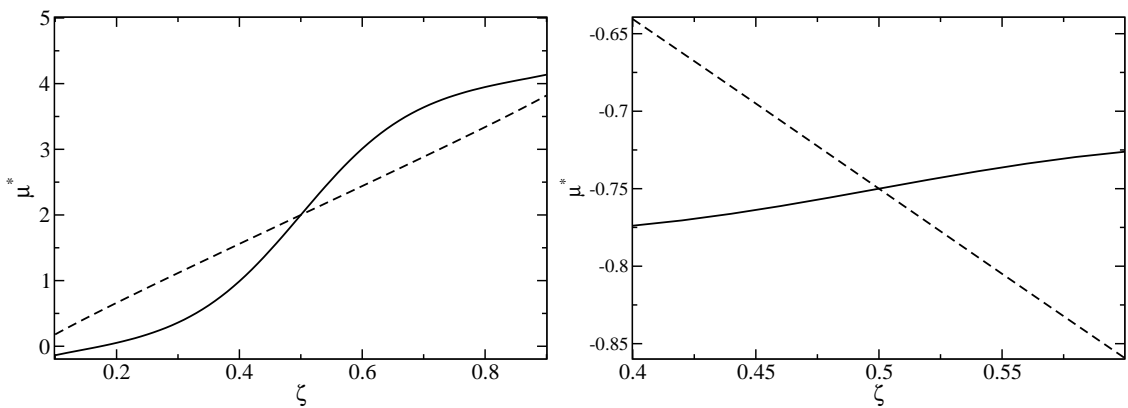

\vskip1cm
\label{mumuMF0_1}
 \includegraphics[scale=0.3]{fig8a.eps}
  \includegraphics[scale=0.3]{fig8b.eps}
 \caption{The reduced chemical potential $\mu^*$  as a function of $\zeta$  (dimensionless) at $T^*=0.1$ for $J=3$ (left panel) 
 and $J=1/4$ (right panel). 
 Solid and dashed lines show Eq. (\ref{mu1d}) 
with and without  the fluctuation contribution (last term) respectively. 
 A narrow range of $\zeta$ is shown for $J=1/4$ because beyond the shown interval there are no solutions of (\ref{cr})-(\ref{vr}).
 $\mu^*$ is in units of $J_1$, with $J_1$ denoting the nearest-neighbour
   attraction.}
   \vskip1cm
\end{figure}

\begin{figure}
\vskip1cm
\label{mumuMF}
\includegraphics[scale=0.33]{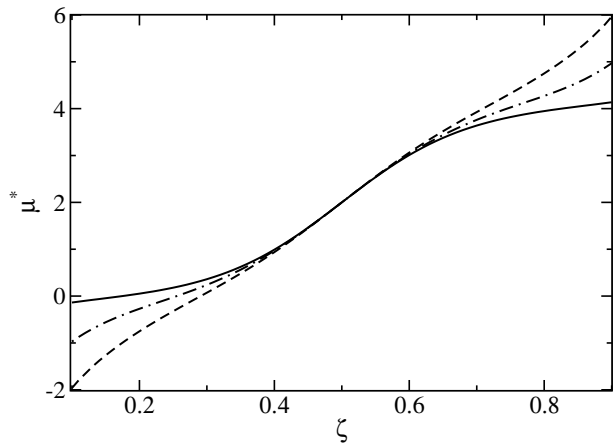}
\caption{ The reduced chemical potential $\mu^*$, Eq.(\ref{muhom}), (in units of $J_1$ with $J_1$ denoting the nearest-neighbor
   attraction)
 for $J=3$ as a function
of the volume fraction $\zeta$ (dimensionless)
deep inside the MF stability region of the periodic phase.
Dashed, dash-dotted and solid lines correspond to  $T^*=0.7,0.4$ and $0.1$ respectively.
}
\vskip1cm
\end{figure}

 When the fluctuation contribution (the last term in Eq. (\ref{mu1d})) is neglected, the slope of the line $\mu^*(\zeta)$ 
 at  low $T^*$ is determined by $\tilde V^*(0)$. The system energy  for $\bar\zeta=const$, 
 $\tilde V(0)$, is positive for 
 the strong repulsion to attraction ratio $J=3$, and negative for $J=1/4$ (see Fig.3),
 therefore in MF the slopes of the $\mu^*(\zeta)$ 
 line are positive and negative
 for $J=3$ and $J=1/4$ respectively. In the former case 3-particle clusters separated by 3 empty sites are 
energetically favourable and no phase separation occurs at $T^*=0$,
whereas in the latter case a separation into dilute and dense phases  occurs at $T^*=0$ 
for $ \mu^*=-0.75$ \cite{pekalski:13:0}.  The MF instability for $J=1/4$ (negative slope of the $\mu^*(\zeta)$ line) is associated with
the MF phase separation that in exact results 
 is absent for $T^*>0$.
For $T^*>0$
 exact results \cite{pekalski:13:0} show positive slopes of $\mu^*(\zeta)$ in both cases,
 in agreements with the results of our theory (solid lines in Fig.8). From the exact results
 it follows that for $J=3$ the slope at $\zeta\approx 1/2$ is very large and increases for decreasing $T^*$, 
 whereas for $J=1/4$ the slope  
 at $\zeta\approx 1/2$ is very small and decreases for decreasing $T^*$. 
 In  Fig.8
 we show the effects of mesoscopic fluctuations on $\mu^*(\zeta)$ for $J=3$ and $J=1/4$ at $T^*=0.1$. The 
very large  difference between the slopes of the $\mu^*(\zeta)$ lines at $\zeta=1/2$ for $J=3$ and $J=1/4$ 
 agrees with exact results. 
 The compressibility at $\zeta=1/2$ is very small for $J=3$ and very large for $J=1/4$;
 the ratio between the compressibility is $\sim 40$.
 
Let us  describe the case of $J=3$ in more detail.
The slope of the $\mu^*(\zeta)$ lines  for $0.4\le \zeta\le 0.6$ is almost independent of $T^*$. Thus, 
our approximate theory does not reproduce the 
decreasing compressibility for decreasing temperature for $\zeta\approx 1/2$.
On the other hand, when $\zeta\le 0.3$ and $\zeta\ge 0.7$ the
decreasing slopes of the $\mu^*(\zeta)$ lines with decreasing temperature 
 are correctly predicted (see Fig.9). 
This behavior indicates that the compressibility increases to very large values
at very small $T^*$ for small and large volume fractions. The increasing compressibility
for decreasing $T^*$ signals the approach to the phase transitions
 that occur at $T^*=0$.
The effect of the last term in Eq.(\ref{muhom}) is clearly seen in Fig.8a.  For $J=3$ this term leads to a 
much smaller slope of the $\mu^*(\zeta)$ line for large- and small volume-fractions, 
and to a much larger slope for the volume fractions 
$\sim 1/2$, than in the absence of mesoscopic fluctuations.
Note also that since $\mu^*\to -\infty$ and  $\mu^*\to \infty$
for $\zeta\to 0$ and $\zeta \to 1$ respectively,
three inflection points at the line $\mu^*(\zeta)$ must be present at low $T^*$. This is a characteristic feature of the
systems with strong mesoscopic  inhomogeneities.

\begin{figure}
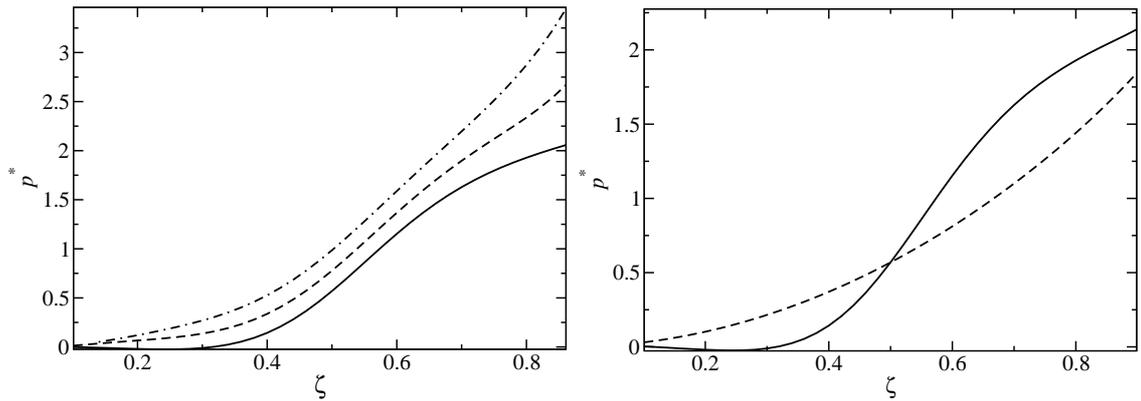

\vskip1cm
  \includegraphics[scale=0.3]{fig10a.eps}
  \includegraphics[scale=0.3]{fig10b.eps}
  \caption{Pressure $p^*$ (in units of the short-range attraction $J_1$ per particle volume)
  as a function of the volume fraction  $\zeta$ (dimensionless) for $J=3$. In the left panel Eq.(\ref{EOS1d}) is shown 
 for  $T^*=0.1$ (solid) 
  $T^*=0.4$ (dash) and $T^*=0.7$  (dash-dotted line). In the right panel the effect of fluctuations is highlighted by
  comparison of the pressure with the MF result (dashed line) for $T^*=0.1$. }
   \vskip1cm
\end{figure}

Most of the above results  agree qualitatively with the exact results in 
the 1d lattice model. On the quantitative level, however, the accuracy of our predictions decreases for decreasing $T^*$.
The exact results for $J=3$ show much smaller compressibility at $\zeta=1/2$ for low $T^*$ than obtained
in our approximation. Moreover, 
the very small compressibility  at $\zeta\approx 1/2$
increases to a  very large value in a  range of $\zeta$ that is much more  narrow than
shown in Fig.8. Finally, the isotherms $\mu^*(\zeta)$ intersect at 3 points: $\mu^*=-2/3,2,14/3$ \cite{pekalski:13:0}, 
whereas in our approximation they are tangent to one another at $\mu^*=2$.

In Fig.10 we present the EOS  for $J=3$. As in the case of the chemical potential, we obtain  qualitative agreement with the
exact results. In particular, at low $T^*$ the slope of the $p^*(\zeta)$ line in the neighborhood of 
$\zeta = 1/2$ is much larger than 
for $\zeta<0.4$ or $\zeta>0.6$. When $T^*$ increases, the   $p^*(\zeta)$ line becomes smoother, in agreement with
Ref.\cite{pekalski:13:0}.
However, the changes of the slope of the $p^*(\zeta)$ line found in Ref.\cite{pekalski:13:0}
are much more pronounced at low $T^*$ than our theory predicts. Our isotherms do not intersect, in contrast to the isotherms
obtained in Ref.\cite{pekalski:13:0}, where they intersect in two points.

\section{discussion}

We have combined DFT and field-theoretic methods in a coarse-grained theory for systems with mesoscopic inhomogeneities
~\cite{ciach:08:1,ciach:11:0,ciach:12:0}. Equations for the average volume fraction 
and the correlation function, (\ref{mu})-(\ref{selfcon}) with (\ref{CG})
have been obtained in the self-consistent Gaussian approximation equivalent to the self-consistent 1-loop 
approximation in the field theory.
Predictions of the theory for the disordered phase, (\ref{muhom})-(\ref{CGF}), 
have been verified by the comparison with the exact results obtained
in Ref.\cite{pekalski:13:0} for a 1d lattice model with first-neighbor attraction and third-neighbor repulsion.
In order to simplify the calculations, we have made further approximations concerning the inverse correlation function in MF, 
Eq.(\ref{tildeCa}), and beyond, Eq. (\ref{tildeCaren}), and adopted for the correlation function in the lattice model 
the expression (\ref{G1d})  obtained for a continuous system. Thus, we have in fact compared a continuous system with the lattice-gas 
form of the entropy, (\ref{fh}), with the exact results obtained for a lattice model.

Despite all the simplifying assumptions, we have obtained a qualitative agreement for all the key features of the system with
mesoscopic inhomogeneities, for the phase-space region where the ansatz (\ref{tildeCaren}) is a reasonable approximation.
First of all, we predict stability of the disordered phase for $T>0$. However, when
the repulsion-to-attraction ratio is large, $J>1$, the disordered phase is strongly inhomogeneous in 
this part of the phase-space region where MF predicts stability of the ordered periodic phase.
In the ordered phase regularly distributed clusters are separated by voids.
In the presence of mesoscopic fluctuations the long-range order is absent, but the inhomogeneity is reflected in the oscillatory
decay of the correlation function. The   correlation length  $\xi_r$  is very large and increases for decreasing $T$.
Both the correlation length and the amplitude of
the correlation function decay rapidly when the MF transition from the periodic to the homogeneous phase is approached 
(Fig.5 for $J=3$). All these features agree with exact results. 

The $\mu(\zeta)$ and EOS isotherms also have the characteristic features observed in Ref. \cite{pekalski:13:0} 
in the case of strong repulsion ($J=3$). 
For low $T$ there are 3 and 2 inflection points in the  $\mu(\zeta)$ and  $p(\zeta)$ curves respectively.
A very large compressibility for the volume fraction $\zeta\approx 0.3$  becomes small
 for $\zeta\approx 0.5$, and very large again for $\zeta\approx 0.7$, in agreement with
Ref.\cite{pekalski:13:0}. 
The very small compressibility results from the repulsion between the clusters when the separation 
between them decreases upon compression.
The very large compressibility accompany the approach to the phase transitions that occur at $T=0$. 
The characteristic shape of the osmotic pressure can be used as an indication of inhomogeneities in experimental studies,
when scattering experiments are not possible. We should note that in the case of weak repulsion, 
when the phase separation between dilute and dense phases 
occurs at $T=0$, the compressibility at $\zeta\approx 0.5$ is very large when $T\to 0$, in a striking 
contrast to the system with the strong repulsion between the particles, $J=3$ (Fig.8). 
The only qualitative feature that is not correctly reproduced by our theory is the decrease of the compressibility
at $\zeta\approx 1/2$ upon a decrease of temperature for 
$J=3$.

On the quantitative level the accuracy
of our results decreases for decreasing $T$.  At very low $T$ we obtain significantly smaller $\xi_r$ and larger
amplitude than in Ref.\cite{pekalski:13:0}.
We should stress that the correlation
function averaged over mesoscopic regions is not expected
to be equal to the standard microscopic correlation
function. Moreover, our numerous approximations applied to simplify the calculations influence the results on the quantitative level. 
Another source of discrepancy between the  self-consistent Gaussian approximation 
and the exact results is the neglected effect of mesoscopic fluctuations on the higher-order correlation functions. 
Finally, the results could be systematically improved within the field-theoretic perturbation expansion beyond the 
self-consistent one-loop approximation. 
Still, the self-consistent Gaussian approximation
gives  results that agree qualitatively with the exact solutions, in contrast to MF,
where a non existing phase transition and a divergent amplitude of the correlation function are obtained.

We conclude that the disordering effect of mesoscopic fluctuations,
neglected in MF, is overestimated in our theory, especially at low $T$. 
We expect significantly better results without the simplifying assumptions, (\ref{tildeCa}), (\ref{tildeCaren}) and 
(\ref{G1d}),
but rather involved numerical 
computations are necessary to solve Eqs.(\ref{muhom})-(\ref{CGF}).

Our results show that most of the qualitative features of the 
 disordered phase with strong inhomogeneities 
 are correctly predicted even when very incomplete information about the interaction potential
is taken into account (see the assumption  (\ref{tildeCa})). Thus, the same properties are obtained at this level of approximation
for all interaction potentials
which in Fourier representation have the same shape near the minimum. More generally, from our theory it follows that 
for $\zeta<\zeta_c$ ( $\zeta>\zeta_c$), where $\zeta_c$ is the critical volume fraction, 
the chemical potential and pressure (see (\ref{muhom}) and (\ref{p})) are smaller (larger)  than predicted by MF. The 
fluctuation correction is the larger the stronger are the inhomogeneities, as measured by $G({\bf r, r})$.

We hope that in future studies
our  self-consistent Gaussian approximation can 
be applied to systems with $d>1$, where the exact results are not feasible. The present comparison with the 
exact results can help to interpret and critically analyze the results.

\acknowledgments
We  gratefully  acknowledge the financial support by the NCN grant 2012/05/B/ST3/03302.

\end{document}